\documentclass[prd,amsmath,amssymb,preprintnumbers,superscriptaddress,twocolumn,10pt,nofootinbib]{revtex4-1} 

\pdfoutput=1
\usepackage{graphicx}
\usepackage{subfigure}
\usepackage{float}

\usepackage{dcolumn}
\usepackage{bm}
\usepackage{amssymb}
\usepackage{latexsym}
\usepackage{booktabs}
\usepackage{amsmath}
\usepackage{multirow}
\usepackage{url}
\usepackage{changes}
\usepackage{soul} 
\usepackage{color, xcolor}
\usepackage[colorlinks=true, linkcolor=red, citecolor=blue]{hyperref}

\hypersetup{
    colorlinks=true,
    linkcolor=red,
    citecolor=blue
}
\IfFileExists{orcidlink.sty}{\usepackage{orcidlink}}{%
  \newcommand{\orcidlink}[1]{%
    \href{https://orcid.org/##1}{%
      \includegraphics[height=1.6ex]{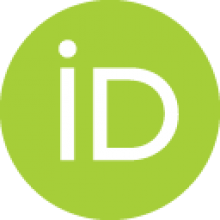}%
    }%
  }%
}

\usepackage[normalem]{ulem}
\usepackage{array}
\usepackage{enumerate}

\def\be{\begin{equation}}
\def\ee{\end{equation}}

\def\bea{\begin{eqnarray}}
\def\eea{\end{eqnarray}}
\newcommand{\bes}{\begin{equation*}}
\newcommand{\ees}{\end{equation*}}
\newcommand{\beqa}{\begin{eqnarray}}
\newcommand{\eeqa}{\end{eqnarray}}

\begin{document}

\title{Synergy between CSST and third-generation gravitational-wave detectors: Inferring cosmological parameters using cross-correlation of dark sirens and galaxies}

\author{Ya-Nan Du\orcidlink{0009-0003-0453-9046}}
\affiliation{Liaoning Key Laboratory of Cosmology and Astrophysics, College of Sciences, Northeastern University, Shenyang 110819, China}

\author{Ji-Yu Song\orcidlink{0009-0003-8111-0470}}
\affiliation{Liaoning Key Laboratory of Cosmology and Astrophysics, College of Sciences, Northeastern University, Shenyang 110819, China}

\author{Yichao Li\orcidlink{0000-0003-1962-2013}}
\affiliation{Liaoning Key Laboratory of Cosmology and Astrophysics, College of Sciences, Northeastern University, Shenyang 110819, China}

\author{Shang-Jie Jin\orcidlink{0000-0003-3697-3501}}
\affiliation{Liaoning Key Laboratory of Cosmology and Astrophysics, College of Sciences, Northeastern University, Shenyang 110819, China}
\affiliation{Department of Physics, University of Western Australia, Perth WA 6009, Australia}
\affiliation{Research Center for the Early Universe, Graduate School of Science, The University of Tokyo, Tokyo 113-0033, Japan}

\author{Ling-Feng Wang\orcidlink{0000-0001-6221-2867}}
\affiliation{School of Physics and Optoelectronic Engineering, Hainan University, Haikou 570228, China}

\author{Jing-Fei Zhang\orcidlink{0000-0002-3512-2804}}
\affiliation{Liaoning Key Laboratory of Cosmology and Astrophysics, College of Sciences, Northeastern University, Shenyang 110819, China}

\author{Xin Zhang\orcidlink{0000-0002-6029-1933}}\thanks{Corresponding author: zhangxin@neu.edu.cn}
\affiliation{Liaoning Key Laboratory of Cosmology and Astrophysics, College of Sciences, Northeastern University, Shenyang 110819, China}
\affiliation{MOE Key Laboratory of Data Analytics and Optimization for Smart Industry, Northeastern University, Shenyang 110819, China}
\affiliation{National Frontiers Science Center for Industrial Intelligence and Systems Optimization, Northeastern University, Shenyang 110819, China}

\begin{abstract}

Gravitational-wave (GW) events are generally believed to originate in galaxies and can thus serve, like galaxies, as tracers of the universe's large-scale structure. In GW observations, waveform analysis provides direct measurements of luminosity distances; however, without relying on a specific cosmological model, the redshifts of GW sources cannot be determined due to the mass-redshift degeneracy. By cross-correlating GW events with galaxies, one can establish a correspondence between luminosity distance and redshift shells, enabling cosmological inference. In this work, we explore the scientific potential of cross-correlating GW sources detected by third-generation (3G) ground-based GW detectors with the photometric redshift survey of the China Space Station Survey Telescope (CSST). We find that the constraint precisions of the Hubble constant and the matter density parameter can reach $1.04\%$ and $2.04\%$, respectively. Additionally, we have also constrained the precision of the GW clustering bias parameter. These results highlight the significant potential of the synergy between CSST and 3G ground-based GW detectors in constraining cosmological models and probing GW source formation channels using cross-correlation of dark sirens and galaxies. 

\end{abstract}

\keywords{Observational cosmology, Black holes, large-scale structure of the Universe, Gravitational wave detectors and experiments, sky surveys}

\pacs{98.80.Es, 97.60.Lf, 98.65.-r, 04.80.Nn, 95.80.+p}

\maketitle

\section{Introduction}\label{sec:Introduction}

Exploring the distribution of the three-dimensional large-scale structure (LSS) of the universe is one of the central goals of modern cosmology. Current optical survey projects, such as the Dark Energy Spectroscopic Instrument (DESI) \cite{DESI:2016fyo}, Euclid \cite{EUCLID:2011zbd,Euclid:2024yrr}, the Vera Rubin Observatory \cite{LSSTScience:2009jmu}, and the China Space Station Survey Telescope (CSST) \cite{CSST:2025ssq}, aim to map the distribution of tens of millions to billions of galaxies across a wide redshift range. By employing statistical tools such as power spectra \cite{SDSS:2001afu,2dFGRS:2005yhx}, correlation functions \cite{SDSS:2005xqv,SDSS:2009kmd}, and tomographic angular power spectra \cite{Bonvin:2011bg,DiDio:2013sea}, these observations can precisely characterize LSS, thereby providing high-precision data to explore the nature of dark energy and dark matter, constrain the cosmic expansion history, and test the theories of gravity \cite{Kaiser:1987qv,Szalay:1997cc,Bonvin:2011bg,Challinor:2011bk,Yoo:2009au}. Recently, DESI has achieved unprecedented precision in constraining the dark energy equation of state by precisely measuring baryon acoustic oscillation (BAO) signals across seven redshift bins and combining them with cosmic microwave background and Type Ia supernova data. The results suggest a 2.8--4.2$\sigma$ preference for a dynamical dark energy evolution \cite{DESI:2025zgx}, sparking wide discussions \cite{Wu:2024faw,Li:2024qso,Du:2024pai,Li:2024qus,Ye:2025ark,Pang:2025lvh,Wu:2025wyk,Li:2025owk,Du:2025iow,Feng:2025mlo,Ling:2025lmw,Li:2025eqh,Li:2025dwz,Li:2025htp,Du:2025xes,Wu:2025vfs,Zhou:2025nkb,Zhang:2025dwu,Li:2025muv}. As China's next-generation space survey telescope, CSST will cover wavelengths from the near-ultraviolet to the near-infrared, equipped with seven photometric filters and three slitless spectroscopic filters, planning to survey an area of 17500 ${\rm deg}^2$ with redshift coverage extending up to $z\sim4$ \cite{CSST:2025ssq}. A large number of studies have explored that CSST could achieve promising results in galaxy clustering \cite{Zhou:2025iiu, Song:2024esa, Xiong:2024dtx, Lin:2023yso}, weak gravitational lensing \cite{Xiong:2024dtx, Lin:2023yso, Liu:2023qbf}, BAO \cite{Shi:2025hsx, Yan:2024jwz}, and cosmic shear measurements \cite{Gong:2019yxt, Su:2025zuc, Cao:2023icw, Lin:2022aro}.

Complementary to optical surveys, gravitational-wave (GW) observations have opened new windows for cosmology, astronomy, and fundamental physics. Recently, ref.~\cite{KAGRA:2025oiz} tested Hawking's area law and the Kerr nature of black holes using a binary black hole (BBH) GW250114, proving that GW detections can provide valuable observational evidence for the theoretical researches on black hole thermodynamics \cite{Kubiznak:2012wp, Wei:2019yvs, Cai:2024tyv, Xu:2025jrk,Zhao:2025ecg,Zhang:2025cdx}. Since LIGO's first detection of GWs in 2015, the LIGO-Virgo-KAGRA (LVK) GW detector network has detected more than 300 GW events, all originating from the mergers of compact binary coalescences, including BBHs, neutron star-black holes (NSBHs), and binary neutron stars (BNSs). The amplitudes of these GW signals scale inversely with luminosity distances, enabling direct measurement of the luminosity distances of GW sources through GW waveform analysis. However, without assuming a specific cosmological model, the degeneracy between the mass and the redshift in the GW waveform makes it impossible to determine the redshift from waveform analysis alone. Theoretically, a small fraction of GW events could have electromagnetic (EM) counterparts, allowing us to identify the host galaxy and thereby measure their redshifts, called bright sirens \cite{Schutz:1986gp, Zhao:2010sz, Cai:2016sby, Wang:2018lun, Zhang:2018byx, Zhang:2019ylr, Zhang:2019loq, Wang:2019tto, Jin:2020hmc, Qi:2021iic, Jin:2021pcv, Wang:2021srv, Wu:2022dgy, Jin:2022tdf, Wang:2022oou, Jin:2023tou, Yu:2023ico, Jin:2023sfc, Han:2023exn, Jin:2023zhi, Feng:2024mfx, Feng:2024lzh, Han:2024sxm, Han:2025fii, Feng:2025wbz}. Ref.~\cite{Jin:2025dvf} provides an up-to-date review of using GW standard sirens to constrain cosmological parameters. To date, only one bright siren, GW170817, has been observed \cite{LIGOScientific:2017adf, LIGOScientific:2017vwq, LIGOScientific:2017zic, Cai:2017aea}. The majority of GW events are dark sirens, lacking EM counterparts. Even in the era of the third-generation (3G) ground-based GW detectors, bright sirens could constitute less than 1\% of all detections \cite{Han:2023exn,Han:2024sxm,Han:2025fii}. Therefore, developing methods to infer the redshifts of dark sirens is of great importance. Currently, the main approach is to search for potential host galaxies in galaxy catalogs and to infer redshift distributions by combining with population models of GW sources \cite{Chen:2017rfc, Gray:2019ksv, Yu:2020vyy, Zhu:2021aat, Zhu:2021bpp, LIGOScientific:2021aug, Mastrogiovanni:2021wsd, Song:2022siz, Jin:2022qnj, Jin:2023sfc, Jin:2023tou, Li:2023gtu, Yun:2023ygz, Mastrogiovanni:2023emh, Mastrogiovanni:2023zbw, Dong:2024bvw, Xiao:2024nmi, Zhu:2024qpp, Zheng:2024mbo, Song:2025ddm, Dong:2025ikq, Zhang:2025yhi}. Nevertheless, such methods are limited by the incompleteness of galaxy catalogs, assumptions about the population distributions of GW sources, and uncertain galaxy luminosity weightings. Hence, exploring additional pathways for redshift determination is crucial to fully realize the scientific potential of dark sirens.  

In the era of 3G ground-based GW detectors, probing LSS with GW detections will become feasible. Detectors such as the Einstein Telescope (ET) and Cosmic Explorer (CE) could detect millions of GW events within $\sim$10 years of observation, with localization precision reaching to the arcminute level \cite{Song:2022siz,Muttoni:2023prw,Abac:2025saz}, enabling GW sources to be tracers of LSS. Furthermore, in cross-correlation analyses between GW sources and galaxies, GW detections provide luminosity distances, while galaxy surveys directly measure redshifts; the two must transform to each other through the distance-redshift relation. Only the correct distance-redshift relation can yield the strongest cross-correlation signal, thereby constraining cosmological parameters and testing theories of gravity \cite{Oguri:2016dgk, Nair:2018ign, Mukherjee:2019wcg, Bera:2020jhx, Diaz:2021pem, Ghosh:2023ksl}. The feasibility of this approach is based on the expectation that GW sources, including BBHs, BNSs, and NSBHs, occur within galaxies, making GW detections analogous to galaxy surveys in tracing the same underlying matter density field \cite{Bosi:2023amu, Jenkins:2018uac, Jenkins:2018kxc, Jenkins:2019uzp, Bertacca:2019fnt}. Unlike conventional methods that search for dark siren host galaxies event-by-event, this approach is largely immune to catalog incompleteness and population-model assumptions, offering a significant complementary method of determining the redshifts of dark sirens.

In this work, we notice the substantial potential of CSST in cross-correlation studies with GW detections, owing to its wide survey area, deep redshift coverage, and high-precision redshift measurements. We therefore wish to investigate the role played by cross-correlation analyses between the galaxies observed by CSST and GW events detected by 3G ground-based GW detectors in cosmological research. We simulate the observation of BBHs by 3G ground-based GW detectors based on population distribution models inferred from the LVK O1--O3 GW data, and combine them with a mock CSST photometric survey galaxy catalog to perform angular power spectrum auto-correlation and cross-correlation analyses. We estimate the constraint precisions of cosmological parameters and GW clustering bias parameters jointly with the Fisher information matrix (FIM) approach.

This paper is organized as follows. Sect.~\ref{sec:method} outlines the methods of cross-correlation analysis of GW events and galaxies. Sect.~\ref{sec:simulation} outlines the methods of modeling the redshift distribution for the CSST photometric survey and simulating the observational data for 3G ground-based GW detectors. In sect.~\ref{sec:results}, we present the results and make relevant discussions. Finally, sect.~\ref{sec:conclusions} summarizes the key findings and concludes the study.

\section{Methods}\label{sec:method}

In this paper, we use angular power spectra to analyze the clustering of galaxies and GW sources, and in this section, we briefly introduce the relevant definitions and analysis methods, primarily following the formulas and notation conventions presented in ref.~\cite{Pedrotti:2025tfg}. We adopt the publicly available code \texttt{pyccl}\footnote{\url{https://ccl.readthedocs.io/en/latest/}} \cite{LSSTDarkEnergyScience:2018yem} to conduct clustering analysis with galaxies and GW sources. We primarily use number counts to generate tracers and apply the Limber approximation in angular power spectrum calculations. In our analysis, GW sources are treated as tracers in the luminosity-distance space, analogous to galaxies in redshift space. Since \texttt{pyccl} supports redshift-based tracer calculations, we map the GW event distribution from luminosity distance to redshift space using the $\Lambda$CDM model with the parameters fixed to the Planck 2018 results.

\subsection{Angular power spectra of galaxies and GW sources}\label{ssec:multi-tracer}

Galaxies or GW sources can serve as tracers of underlying dark matter, and the distribution of their number density fluctuations reflects the distribution of underlying dark matter. For a given tracer $X$ (galaxies or GW sources), the number density contrast is expressed as
\begin{equation}\label{eq:1}
\Delta^X(\boldsymbol{n},x)=\frac{N^X(\boldsymbol{n},x)-\langle N^X\rangle(x)}{\langle N^X\rangle(x)},   
\end{equation}
where $N^{X}(\boldsymbol{n},x)$ is the number count at radial position $x$ in the direction $\boldsymbol{n}$, and $\langle...\rangle$ denotes the ensemble average, which is an average over all directions.

We can further expand the number density fluctuation with spherical harmonic functions as follows:
\begin{equation}\label{eq:2}
\Delta^X(\boldsymbol{n},x)=\sum_{\ell=0}^\infty\sum_{m=-\ell}^\ell s_{\ell m}^X(x)Y_{\ell m}(\boldsymbol{n}),
\end{equation}
where $s_{\ell m}^X(x)$ is the spherical harmonic coefficient and $Y_{\ell m}(\boldsymbol{n})$ is the corresponding spherical harmonic function. 

To obtain the statistic properties of the underlying dark matter field, we divide galaxies in redshift space and GW sources in luminosity distance space into several bins, and compute the tomographic angular power spectrum between the $i$-th bin and $j$-th bin, $C_{\ell}^{XY}(x_{i},x_{j})$, by cross-correlating the harmonic coefficients of different tracers in these bins, given by
\begin{equation}\label{eq:3}
\langle s_{\ell m}^{X}(x_{i})s_{\ell^{\prime}m^{\prime}}^{Y^{*}}(x_{j})\rangle=\delta_{\ell\ell^{\prime}}\delta_{mm^{\prime}}C_{\ell}^{XY}(x_{i},x_{j}),
\end{equation}
where ${^*}$ denotes the complex conjugate; $X$ and $Y$ represent two tracers, with the case of $X=Y$ correspondind to the auto-correlation and that of otherwise representing the cross-correlation; $\delta$ is the Kronecker delta.

Specifically, the angular power spectrum is derived from the expression of the tracer number density fluctuations in perturbation theory, where the fluctuations are weighted and binned in redshift or luminosity distance intervals as given by
\begin{equation}
C_\ell^{XY}(x_i,x_j)=\frac{2}{\pi}\int\mathrm{d}kk^2P(k)\Delta_\ell^{X,x_i}(k)\Delta_\ell^{Y,x_j}(k),
\end{equation}\label{eq:4}
where $P(k)$ denotes the primordial power spectrum, and $\Delta_\ell^{X,x_i}(k)$ is the effective Fourier transfer function of the tracer $X$ in the $i$-th bin, given by
\begin{equation}\label{eq:5}
\Delta_\ell^{X,x_i}(k)=\int_0^\infty\mathrm{d}xw^X(x,x_i)\Delta_\ell^X(k,x),
\end{equation}
where $\Delta_\ell^X(k,x)$ is the number density fluctuation transfer function in the Fourier domain, containing contributions from density, velocity, lensing, and gravitational effects \cite{Bonvin:2011bg,Challinor:2011bk,Fonseca:2023uay}. The velocity effects further comprise redshift-space distortions (RSDs) and Doppler terms \cite{Saga:2022frj, vanderSteen:2025nxy}. In the luminosity distance space, RSD corresponds to the luminosity distance space distortion (LSD) \cite{Zhang:2018nea,Vijaykumar:2020pzn,Libanore:2020fim}. Consequently, $\Delta_\ell^X(k,x)$ is written as
\begin{equation}\label{eq:6}
\begin{aligned}
    \Delta_{\ell}^{X}(k,x)=&\Delta_\ell^{X,\mathrm{den}}(k,x)+\Delta_\ell^{X,\mathrm{vel}}(k,x)\\&+\Delta_\ell^{X,\mathrm{len}}(k,x)+\Delta_\ell^{X,\mathrm{gr}}(k,x),
\end{aligned}
\end{equation}
where $\Delta_\ell^{X,\mathrm{den}}(k,x)$, $\Delta_\ell^{X,\mathrm{vel}}(k,x)$, $\Delta_\ell^{X,\mathrm{len}}(k,x)$, and $\Delta_\ell^{X,\mathrm{gr}}(k,x)$ represent the contributions from the density, velocity, lensing, and gravitational effects, respectively. The velocity term is further given by
\begin{equation}\label{eq:7}
\Delta_\ell^{X,\mathrm{vel}}(k,x)=\Delta_\ell^{X,\mathrm{RSD/LSD}}(k,x)+\Delta_\ell^{X,\mathrm{dop}}(k,x),
\end{equation}
with the two components corresponding to the RSD/LSD and Doppler effects, respectively.

Following ref.~\cite{Pedrotti:2025tfg}, we only consider the dominant contributions to the signal, namely the density term, lensing term, and the RSD/LSD term.

The term $w^{X}(x,x_{i})$ in eq.~\eqref{eq:5} is a weighted window function centered at coordinate $x_i$, which is defined by the unweighted window function $W^{X}(x,x_{i})$, as follows:
\begin{equation}\label{eq:8}
w^{X}(x,x_{i})=W^{X}(x,x_{i})\frac{\mathrm{d}N_{\mathrm{obs}}^{X}}{\mathrm{d}x}\frac{1}{\int\mathrm{d}x^{\prime}W^{X}(x^{\prime},x_{i})\frac{\mathrm{d}N_{\mathrm{obs}}^{X}}{\mathrm{d}x^{\prime}}}.
\end{equation}

The unweighted window function $W^{X}(x,x_{i})$ is expressed as the convolution of a step function $S^{X}(x,x_{i})$ with a log-normal likelihood $\mathcal{L}(x^{obs}|x)$ encoding measurement errors, given by
\begin{equation}\label{eq:9}
\begin{aligned}
W^X(x,x_i) & =\int_0^\infty\operatorname{d}yS^X(y,x_i)\mathcal{L}(y|x) \\
           & =\frac{1}{2}\left\{\operatorname{erf}[u(x^{i+1},x)]-\operatorname{erf}[u(x^i,x)]\right\},
\end{aligned}
\end{equation}
where erf is the error function, and $u$ is given by
\begin{equation}\label{eq:10}
u(y,x)=\frac{\ln y-\ln x}{\sqrt{2}\sigma_{\ln x}},
\end{equation}
where $\sigma_{\ln x}$ is the standard deviation of the log-normal distribution, corresponding to the relative error in the redshift/luminosity distance measurements. $x^i$ and $x^{i+1}$ represent the lower and upper boundaries of the $i$-th bin, respectively.

The term $\mathrm{d}N_{\mathrm{obs}}^{X}/\mathrm{d}x^{\prime}$ denotes the observed number density distribution of tracer $X$. The detailed procedures for deriving the observed number densities of both galaxies and GW sources are presented in sect.~\ref{sec:simulation}. This distribution also determines the binning strategy. Specifically, we first divide the CSST photometric sample into 15 redshift bins containing equal numbers of galaxies, and then follow ref.~\cite{Pedrotti:2025tfg}, convert them into 15 luminosity-distance bins based on the Planck 2018 cosmology. These bins are subsequently applied in the GW samples analysis. For completeness, we also explore alternative binning schemes to assess the robustness of our results. For details see sect.~\ref{ssec:factors}.

By adopting the Limber approximation, small-scale fluctuations along the line of sight can be neglected, allowing for a focus on transverse structures. This approach is sufficiently accurate at small angular scales and is well-suited for processing cosmological signals in the linear to mildly nonlinear regimes \cite{Gao:2023tcd}. After applying the Limber approximation to eq.~\eqref{eq:4}, the GW-galaxy cross-correlation, the galaxy auto-correlation, and the GW auto-correlation can be simplified. In our calculations, we include the contributions from density, lensing, and RSD/LSD terms. For their detailed expressions, we refer the readers to Appendix A of ref.~\cite{Pedrotti:2025tfg}. Here, we present only the dominant density-density terms for brevity:
\begin{equation}\label{eq:11}
\begin{aligned}
C_{{\ell},\mathrm{den}}^{\mathrm{gGW}}(z_{i},d_{\mathrm{L},j})
=& \int_{0}^{\infty}\mathrm{d}z\,w^{\mathrm{g}}(z,z_{i})\,w^{\mathrm{GW}}[d_{\rm{L}}(z,\lambda),d_{\mathrm{L},j}]\\
&\times\frac{\mathrm{d}d_{\mathrm{L}}}{\mathrm{d}z}(z,\lambda) \frac{H(z)}{cr(z)^{2}}b_{\mathrm{GW}}(z)b_{\mathrm{g}}(z)\\
&\times\mathcal{P}\left[\frac{\ell+1/2}{r(z,\lambda)},z\right],
\end{aligned}
\end{equation}
\begin{equation}
\begin{aligned}
C_{{\ell},\mathrm{den}}^{\mathrm{gg}} (z_{i},z_{j}) 
=& \int_{0}^{\infty}\mathrm{d}z\,w^{\mathrm{g}}(z,z_{i})\,w^{\mathrm{g}}(z,z_{j})\frac{H(z)}{cr(z)^{2}}\\
&\times\left[b_{\mathrm{g}}(z)\right]^2\,\mathcal{P}\left[\frac{\ell+1/2}{r(z,\lambda)},z\right],
\end{aligned}
\end{equation}
\begin{equation}
\begin{aligned}
C_{{\ell},\mathrm{den}}^{\mathrm{GWGW}}(d_{\mathrm{L},i},d_{\mathrm{L},j})
=& \int_{0}^{\infty}\mathrm{d}z\,w^{\mathrm{GW}}[d_{\rm{L}}(z,\lambda),d_{\mathrm{L},i}]\\
&\times{w^{\mathrm{GW}}[d_{\rm{L}}(z,\lambda),d_{\mathrm{L},j}]}\\
&\times\left[\frac{\mathrm{d}d_{\mathrm{L}}}{\mathrm{d}z}(z,\lambda)\right]^2\frac{H(z)}{cr(z)^{2}}\\
&\times\left[b_{\mathrm{GW}}(z)\right]^2\,\mathcal{P}\left[\frac{\ell+1/2}{r(z,\lambda)},z\right],
\end{aligned}
\end{equation}
where $b_{\mathrm{g}}$ and $b_{\mathrm{GW}}$ represent bias parameters of galaxies and GW sources, respectively; ${r(z,\lambda)}$ is the comoving distance, and $\lambda$ represents cosmological parameters in the distance-redshift relation. 
The weighting functions in eq.~\eqref{eq:4} exhibit distinct dependencies: $w^{\mathrm{g}}$ is defined in redshift space, whereas $w^{\mathrm{GW}}$ is defined in luminosity distance space. For a given redshift bin $z_i$ where $w^{\mathrm{g}}$ has non-zero support, the corresponding GW window function $w^{\mathrm{GW}}$ will contribute significantly to the angular power spectrum only if the distance bin $d_{\mathrm{L},j}$ encompasses the range of $d_{\rm L}(z, \Omega)$ mapped from $z_i$ under the assumed cosmology. The cross-correlation signal $C_{{\ell}}^{\mathrm{gGW}}$ is maximized when these two window functions achieve optimal overlap in the radial integration \cite{Pedrotti:2025tfg}. Since any deviation from the true fiducial cosmology would shift the relative alignment between the redshift and distance bins, the cross-correlation amplitude is highly sensitive to the underlying expansion history. Consequently, the strength of the cross-correlation signal can be utilized as a powerful probe to constrain the distance-redshift relation. The intrinsic properties of tracers can also affect the cross-correlation signal in different ways. In this work, for simplicity, we consider only three factors: the observed redshift/luminosity distance distribution, the clustering bias, and the magnification bias.

The term $b_{X}$ represents the clustering bias of tracer $X$. For any tracer $X$, its density term connects to the dark matter density field through a bias term $b_{X}=\delta_{X}/\delta_{\rm m}$. Within the large-scale regime considered in this work, this relationship can be approximated as being dependent solely on redshift and independent of scale. For the clustering bias of the CSST galaxy survey, we adopt the parameterization from ref.~\cite{Gong:2019yxt}, expressed as:
\begin{equation}
b_{\mathrm{g}}(z) = b_{0}(1+z)^{b_{1}},
\end{equation}
where $b_{0}$ and $b_{1}$ are free parameters. In this work, we fix them to $b_{0}=1$ and $b_{1}=1$ for simplicity \cite{Gong:2019yxt}. For the GW survey, the clustering bias follows the same functional form, given by
\begin{equation}\label{eq:GW bias}
    b_{\mathrm{GW}}(z)=A_{\mathrm{GW}}(1+z)^{\gamma}.
\end{equation}
In this work, we adopt $A_{\mathrm{GW}} = 1.20$ and $\gamma = 0.59$ as fiducial values \cite{Peron:2023zae}, while simultaneously constraining both the GW clustering bias parameters and the cosmological parameters.

Similar to eq.~\eqref{eq:11}, the lensing cross-term in eq.~\eqref{eq:4} under the Limber approximation introduces a magnification bias term $s_{X}$ (see Appendix A of the ref.~\cite{Pedrotti:2025tfg} for details). This bias primarily quantifies the change in the observed number of sources due to gravitational lensing effects. For the specific treatment of magnification bias in both galaxy survey and GW survey, we refer to ref.~\cite{Pedrotti:2025tfg}: $s(z)=s_{0}+s_{1}z+s_{2}z^{2}+s_{3}z^{3}$, where we adopt the relevant data $s_{0}=0.0842$, $s_{1}=0.0532$, $s_{2}=0.298$, $s_{3}=-0.0113$ for the galaxy survey, and for the GW survey, $s_{0}=-0.00559$, $s_{1}=0.0292$, $s_{2}=0.00344$, $s_{3}=0.00258$. In addition to affecting lensing effects, magnification bias also influences the gravitational and Doppler terms \cite{Pedrotti:2025tfg, Bonvin:2016dze}.

It should be noted that we have adopted the Limber approximation in calculating the angular power spectra. While this approximation is sufficiently accurate at small angular scales, it is not precise on large scales \cite{Bellomo:2020pnw, Bernal:2020pwq}.

\subsection{Noise and the range of angular scales}\label{ssec:noise}

\begin{table*}
\centering
\caption{Bin edges and corresponding values for the minimum multipole $\ell_{\min}$, the maximum multipole $\ell_{\max}^{\mathrm{NL}}$ adopted, based on restricting the analysis to linear/mildly nonlinear scales, and values of the best resolved angular scale $\ell_{\max}^{\mathrm{LOC}}$ and damping scale $\ell_{\mathrm{damp}}$, coming from the finite sky resolution of GW events. This table refers to the ET2CE configuration.}\label{tab:ell}
\centering
\renewcommand{\arraystretch}{2}
\begin{tabular}{ccccccc}
\hline\hline 
\makebox[0.17\textwidth][c]{Bin Edges} & \makebox[0.17\textwidth][c]{$\ell_{\min}$} & \makebox[0.17\textwidth][c]{$\ell_{\max}^{\mathrm{NL}}$} & \makebox[0.17\textwidth][c]{$\ell_{\max}^{\mathrm{LOC}}$} &
\makebox[0.17\textwidth][c]{$\ell_{\mathrm{damp}}$}\\
\hline
0 -- 0.25 & 5 & 40 & 4204 & 907 \\
0.25 -- 0.35 & 5 & 103 & 1331 & 489 \\
0.35 -- 0.43	& 5	& 139 & 1173 & 372 \\
0.43 -- 0.51	& 5	& 173 & 872 & 307 \\ 
0.51 -- 0.58	& 10 & 207 & 757 & 250 \\
0.58 -- 0.65	& 10 & 240 & 563 & 205 \\ 
0.65 -- 0.73	& 10 & 278 & 512 & 172 \\ 
0.73 -- 0.81	& 10 & 319 & 374 & 161 \\
0.81 -- 0.90	& 15 & 366 & 364 & 130 \\
0.90 -- 1.00	& 15 & 421 & 278 & 109 \\
1.00 -- 1.11	& 15 & 486 & 256 & 99 \\
1.11 -- 1.25	& 15 & 569 & 244 & 89 \\
1.25 -- 1.44	& 20 & 587 & 189 & 76 \\
1.44 -- 1.74	& 20 & 884 & 152 & 61 \\
1.74 -- 4.00	& 20 & 2386 & 99 & 40 \\
\hline\hline
\end{tabular}
\end{table*}

\begin{figure}
    \centering
    \includegraphics[width=1\linewidth]{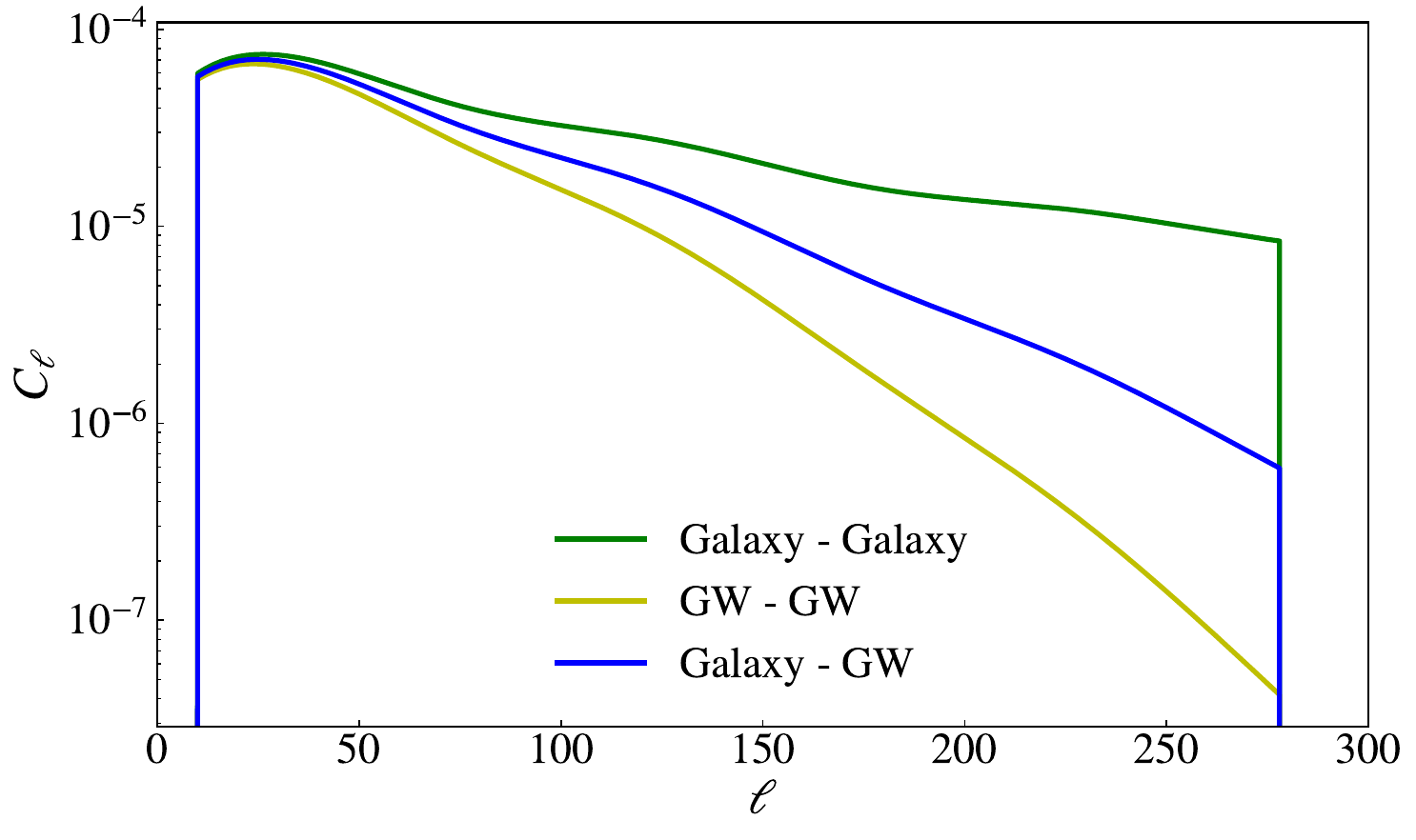}
    \includegraphics[width=1\linewidth]{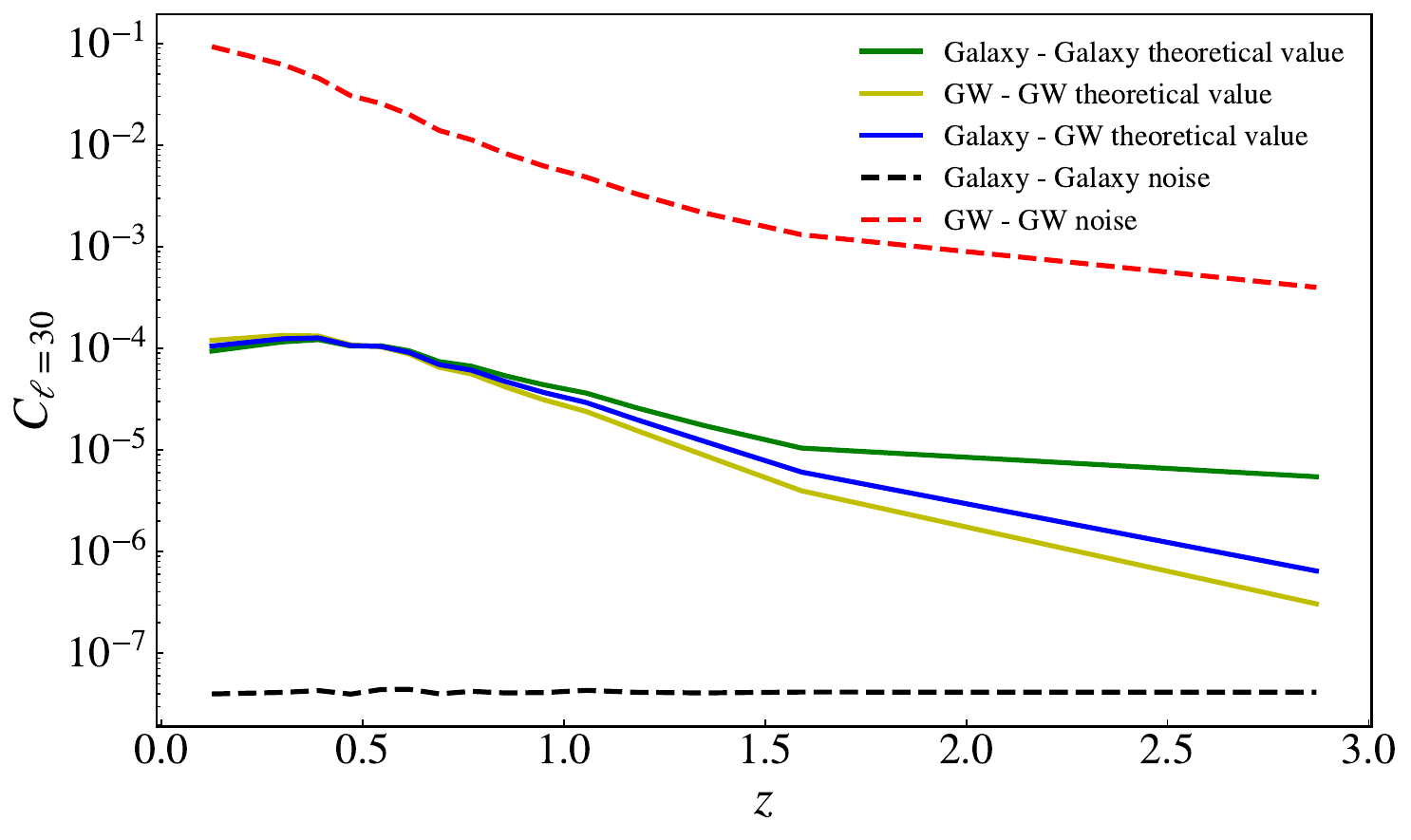}
    \caption{Angular power spectra and noise contributions. The upper panel shows the dependence on multipoles $\ell$ in the redshift bin $z\in[0.65,0.73]$, while the lower panel presents the redshift evolution at $\ell=30$. Green, yellow, and blue solid curves correspond to the galaxy auto-spectrum, GW auto-spectrum, and galaxy-GW cross-spectrum, respectively. Black and red dashed lines indicate the noise in the galaxy and GW auto-spectra, respectively.}
    \label{fig:clnl}
\end{figure}

We supplement the theoretical angular power spectrum with the corresponding noise power spectrum. By expanding the noise component into spherical harmonic coefficients $n_{\ell m}(x_i)$, the total harmonic coefficients of the observed signal can be written as follows:
\begin{equation}\label{eq:12}
a_{\ell m}^X(x_i)=s_{\ell m}(x_i)+n_{\ell m}(x_i).
\end{equation}

The expectation value of the noise is given by
\begin{equation}\label{eq:13}
\langle n_{\ell m}^X(x_i)n_{\ell^{\prime}m^{\prime}}^{Y^*}(x_j)\rangle=\delta_{\ell\ell^{\prime}}\delta_{mm^{\prime}}\delta_{XY}\delta_{ij}\mathcal{N}^X(x_i),
\end{equation}
where $\mathcal{N}^X(x_i)$ is the reciprocal of the total number of sources observed per steradian in the $i$-th bin, expressed as:
\begin{equation}\label{eq:14}
\mathcal{N}^X(x_i)=\left[\int_0^\infty\mathrm{d}xW^X(x,x_i)\frac{\mathrm{d}^2N_{\mathrm{obs}}^X}{\mathrm{d}x\mathrm{d}\Omega}\right]^{-1}.
\end{equation}

Assuming that the signal and noise are uncorrelated, i.e., $\langle n_{\ell m}^X(x_i)s_{\ell^{\prime}m^{\prime}}^{Y^*}(x_j)\rangle = 0$, the observed angular power spectrum can be expressed as:
\begin{equation}\label{eq:15}
\langle a_{\ell m}^X(x_i),a_{\ell^{\prime}m^{\prime}}^{Y^*}(x_j)\rangle=\delta_{\ell\ell^{\prime}}\delta_{mm^{\prime}}\tilde{C}_\ell^{XY}(x_i,x_j),
\end{equation}
\begin{equation}\label{eq:16}
\tilde{C}_\ell^{XY}(x_i,x_j)\equiv C_\ell^{XY}(x_i,x_j)+\delta_{XY}\delta_{ij}\mathcal{N}^X(x_i).
\end{equation}

In addition to the measurement uncertainties in redshift and luminosity distance considered in eq.~\eqref{eq:10}, we also account for the sky localization errors of GW sources. Accordingly, the cross- and auto-angular power spectra, $C_\ell^{\mathrm{gGW}}$ and $C_\ell^{\mathrm{GWGW}}$, are modified as follows:
\begin{equation}\label{eq:17}
C_\ell^{\mathrm{gGW}}(d_{\mathrm{L},i},z_j)\mapsto C_\ell^{\mathrm{gGW}}(d_{\mathrm{L},i},z_j)\times e^{-\ell(\ell+1)/\ell_{\mathrm{damp}}^2},
\end{equation}
\begin{equation}
\begin{aligned}\label{eq:18}
    C_\ell^{\mathrm{GWGW}}(d_{\mathrm{L},i},d_{\mathrm{L},j})\mapsto & C_\ell^{\mathrm{GWGW}}(d_{\mathrm{L},i},d_{\mathrm{L},j})\\&\times e^{-2\ell(\ell+1)/\ell_{\mathrm{damp}}^2},
\end{aligned}
\end{equation}
\begin{equation}\label{eq:19}
\ell_{\mathrm{damp}}^2=\frac{(2\pi)^{3/2}}{\Delta\Omega_{1\sigma}},
\end{equation}
where $\Delta\Omega_{1\sigma}$ is computed from the 50th percentile of the 1${\sigma}$ sky localization uncertainties within each bin, expressed as $\Delta\Omega_{1\sigma}=Q_{50}^{\Delta\Omega}(d_{\mathrm{L},i})$. 

For the determination of the maximum angular scale $\ell_{\max}$, different criteria are applied depending on the type of power spectrum. For the galaxy auto-correlation, $\ell_{\max}$ is limited solely by the nonlinear cutoff, whereas for the GW auto-correlation and GW--galaxy cross-correlation, both the nonlinear scale and the angular-resolution limit of GW localization must be considered, and hence $\ell_{\max}$ is taken as the minimum of the two limits. 

Sky localization errors affect the accessible angular scales of GW sources \cite{Libanore:2020fim}. For each redshift bin, we adopt the 99th percentile of sky localization uncertainties at the 90\% confidence level, denoted by $\Delta\Omega_{\min}$ \cite{Pedrotti:2025tfg}, to define the maximum values of the angular scale limited by the angular resolution, given by 
\begin{equation}\label{eq:20}
    \ell_{\max}^{\mathrm{LOC}} = \pi / \sqrt{\Delta\Omega_{\min}}.
\end{equation}

To ensure that the analysis remains within the linear regime, the maximum values of the angular scale affected by the nonlinear cutoff is defined as:
\begin{equation}\label{eq:21}
    \ell_{\rm max}^{\mathrm{NL}} = r(z_i) k_{i,\mathrm{max}},
\end{equation}
where $r(z_i)$ is the comoving distance of the $i$-th redshift bin and $k_{i,\mathrm{max}} = \pi / (2R_i)$. The characteristic radius $R_i$ is determined by requiring $\sigma^2(R_i) = 0.25$ \cite{Pedrotti:2025tfg}, corresponding to the linear--nonlinear transition scale. According to our calculation, $\ell_{\max}^{\mathrm{NL}}$ increases with redshift, while $\ell_{\max}^{\mathrm{LOC}}$ decreases due to the worsening sky localization at higher redshifts.

For the minimum multipole $\ell_{\min}$, we follow ref.~\cite{Tanidis:2019teo} and adopt the following values \cite{Pedrotti:2025tfg}: $\ell_{\min}=5$ for $z<0.5$, $\ell_{\min}=10$ for $0.5<z<0.75$, $\ell_{\min}=15$ for $0.75<z<1.25$, and $\ell_{\min}=20$ for $z>1.25$. The adopted values of $\ell_{\min}$, $\ell_{\max}$, and the damping scale $\ell_{\mathrm{damp}}$ for each redshift bin are summarized in Table~\ref{tab:ell}.

Based on the determined angular scale range for each redshift bin, the upper panel of Figure~\ref{fig:clnl} presents the theoretical angular power spectrum $C_\ell$ as a function of the multipole $\ell$ in the seventh redshift bin, including contributions from density, lensing and RSD/LSD effects. Due to the damping effect caused by GW sky localization uncertainties, both the GW–GW and galaxy–GW spectra exhibit varying degrees of attenuation. The effect is relatively weak at low $\ell$, but strengthens significantly as the multipole increases, eventually leading to strong attenuation of the signal at high $\ell$. The lower panel displays the redshift evolution of the theoretical angular power spectrum and the noise level of the auto-correlation spectrum at fixed multipole value. It can be seen that the noise in GW observations is notably higher than that in galaxy observations. Additionally, since the density term dominates the theoretical angular power spectrum, the clustering bias parameters have a significant influence on it. At low redshifts, the galaxy bias $b_{\mathrm{g}}(z)$ is smaller than the GW bias $b_{\mathrm{GW}}(z)$, so the galaxy–galaxy spectrum lies below the GW–GW spectrum. As redshift increases, the galaxy bias becomes larger than the GW bias, causing the galaxy–galaxy spectrum to gradually exceed the GW–GW spectrum. This transition specifically occurs in the fifth redshift bin $[0.51, 0.58]$, as can also be seen around redshift $z = 0.5$ in the lower panel. The seventh redshift bin $[0.65, 0.73]$ shown in the upper panel corresponds to the initial stage after this transition is completed. In this bin, the galaxy–galaxy spectrum is only slightly higher than the GW–GW spectrum, and the difference is small. Therefore, at low $\ell$ where the attenuation is weak, the two angular power spectra are very close to each other.

\subsection{Forecasts of cosmological parameter constraints and SNRs}\label{ssec:fisher matrix}

We employ the FIM approach to estimate constraint errors of cosmological parameters. For a given set of cosmological parameters ${\theta_\alpha, \theta_\beta}$, the corresponding elements of FIM is given by
\begin{equation}
\begin{aligned}
    F_{ij}&=-\left\langle\frac{\partial^2\ln\mathcal{L}}{\partial\theta_\alpha\partial\theta_\beta}\right\rangle\\&=f_\mathrm{sky}\sum_\ell\frac{2\ell+1}{2}\mathrm{Tr}[\mathcal{C}_\ell^{-1}(\partial_\alpha\mathcal{C}_\ell)\mathcal{C}_\ell^{-1}(\partial_\beta\mathcal{C}_\ell)],
\end{aligned}
\end{equation}
where $f_{\mathrm{sky}} \equiv \Omega_{\mathrm{sky}} / (41253~\mathrm{deg}^2)$ is the sky coverage fraction, $\Omega_\mathrm{sky}$ represents the sky area covered by the survey observation, measured in square degrees, and the total area of the entire sky is approximately 41253 deg$^2$. $\mathcal{C}_\ell$ is the covariance matrix. Assuming the galaxy survey is divided into $N$ tomographic redshift bins $z_1$, ..., $z_N$, and the GW survey is divided into $M$ luminosity-distance bins $d_{\mathrm{L},1}$, ..., $d_{\mathrm{L},M}$, the symmetric covariance matrix can be expressed as:
\begin{equation}
\begin{aligned}
    &\mathcal{C}_\ell =\\
&\resizebox{\columnwidth}{!}{$
\begin{pmatrix}
\tilde{C}_\ell^{\rm gg}(z_1,z_1) & \cdots & \tilde{C}_\ell^{\rm gg}(z_1,z_N)
 & \tilde{C}_\ell^{\rm gGW}(z_1,d_{\mathrm{L},1}) & \cdots 
 & \tilde{C}_\ell^{\rm gGW}(z_1,d_{\mathrm{L},M}) \\[4pt]
 & \ddots & \vdots & \vdots & & \vdots \\[4pt]
 & & \tilde{C}_\ell^{\rm gg}(z_N,z_N)
 & \tilde{C}_\ell^{\rm gGW}(z_N,d_{\mathrm{L},1}) & \cdots 
 & \tilde{C}_\ell^{\rm gGW}(z_N,d_{\mathrm{L},M}) \\[4pt]
 & & & \tilde{C}_\ell^{\rm GWGW}(d_{\mathrm{L},1},d_{\mathrm{L},1}) 
 & \cdots & \tilde{C}_\ell^{\rm GWGW}(d_{\mathrm{L},1},d_{\mathrm{L},M}) \\[4pt]
 & & & & \ddots & \vdots \\[4pt]
 & & & & & \tilde{C}_\ell^{\rm GWGW}(d_{\mathrm{L},M},d_{\mathrm{L},M})
\end{pmatrix}
$},
\end{aligned}
\end{equation}
where $\tilde{C}_\ell$ is the observed angular power spectrum given in eq.~\eqref{eq:16}. The term $\partial_\alpha\mathcal{C}_\ell=\partial\mathcal{C}_\ell/\partial\theta_\alpha$ contains the derivatives of the covariance matrix elements with respect to the given cosmological parameters, while keeping all the others fixed, specifically expressed as:
\begin{equation}\label{eq:24}
\begin{split}
\frac{\partial\mathcal{C}_\ell}{\partial\theta_\alpha} = \frac{\mathcal{C}_\ell(\theta_\alpha+\Delta\theta_\alpha)-\mathcal{C}_\ell(\theta_\alpha-\Delta\theta_\alpha)}{2\Delta\theta_\alpha},
\end{split}
\end{equation}
where $\Delta\theta_\alpha$ is a small perturbation of the parameter $\theta_\alpha$ around the fiducial value of the parameter $\theta_\alpha$. We define a relative step-size factor $f_\alpha$ for each parameter to determine the absolute step size $\Delta\theta_\alpha = f_\alpha \cdot \theta_\alpha$. The specific values are: $f_h = 1.0\times10^{-6}$, $f_{\Omega_{\mathrm{c}}} = 1.0\times10^{-4}$, $f_{\Omega_{\mathrm{b}}} = 1.0\times10^{-4}$, $f_{n_{\mathrm{s}}} = 1.0\times10^{-5}$, $f_{A_{\mathrm{GW}}} = 1.0\times10^{-4}$, and $f_{\gamma} = 1.0\times10^{-4}$. For the logarithmic parameter $\ln(10^{10}A_{\rm s})$, we directly assign an absolute step size of $\Delta{\ln A_{\rm s}} = 1.0\times10^{-3}$ to ensure numerical stability. To verify the robustness of these choices, we performed a systematic convergence test by varying each step-size factor over one order of magnitude around its chosen value. Our results indicate that the relative changes in the derivatives remain below 1\%, confirming that the chosen step sizes lie within a stable convergent plateau and that numerical errors have a negligible impact on the forecasted cosmological constraints. For a given parameter $\theta_{\alpha}$, the estimate 1$\sigma$ error is given by $\sigma_{\theta_{\alpha}}=\sqrt{F_{\alpha\alpha}^{-1}}$.

Regarding the calculation of the FIM, when we take partial derivatives with respect to cosmological parameters, we recompute the mapping between redshift and luminosity distance for the corresponding parameter values. Unlike ref.~\cite{Pedrotti:2025tfg}, which evaluates cross-correlations in both redshift space and distance space, our analysis uses fixed redshift bins. Galaxy redshift observations directly define the boundaries of these bins, and the cosmological model maps them into luminosity-distance bins. We then assign GW events to these luminosity-distance bins according to their measured luminosity distances. As cosmological parameters change, the boundaries of the luminosity-distance bins also change, which in turn changes the distribution of events among the bins.

To evaluate the detectability of the cross-correlation signal, we compute its signal-to-noise ratio (SNR). Following the approach of ref.~\cite{Scelfo:2020jyw}, we construct a data vector by stacking the theoretical angular power spectra of different tracers and bin combinations as follows:
\begin{equation}\label{eq:24}
\begin{split}
\mathbf{C}_\ell = (&\, C_\ell^{\rm gg}(z_1,z_1),\ \ldots, 
C_\ell^{\rm gGW}(z_1,d_{\mathrm{L},1}),\ \ldots,\\
&\, C_\ell^{\rm GWGW}(d_{\mathrm{L},1},d_{\mathrm{L},1}),\ \ldots )^\top.
\end{split}
\end{equation}

Each element is assigned to a specific pair of tracers $[X_1, X_2]$, denoting the types of tracers and their redshift bins. For example, the first element is associated to the couple of indices [$X_1=g_{z_1}, X_2=g_{z_1}$]. The covariance matrix is defined as follows:
\begin{equation}\label{eq:25}
[\mathrm{Cov}(\ell)]_{XY}=\tilde{C}_{\ell}^{X_{1}Y_{1}}\tilde{C}_{\ell}^{X_{2}Y_{2}}+\tilde{C}_{\ell}^{X_{1}Y_{2}}\tilde{C}_{\ell}^{X_{2}Y_{1}}.
\end{equation}
The diagonal elements of this matrix yield the $1\sigma$ uncertainties on the corresponding parameters:
\begin{equation}
    \begin{aligned}
    \delta C_\ell^{XY}(x_i,x_j)=&\bigg[\frac{1}{(2\ell+1)f_{\mathrm{sky}}} \bigg(C_\ell^{XY}(x_i,x_j)^2+\\ &\tilde{C}_\ell^{XX}(x_i,x_i)\times\tilde{C}_\ell^{YY}(x_j,x_j)\bigg\}\bigg]^{\frac{1}{2}}.
    \end{aligned}
\end{equation}

SNR for a specific tracer pair combination $[X_1, X_2]$ at a given multipole $\ell$ is given by
\begin{equation}\label{eq:27}
\begin{aligned}
\mathrm{SNR}_{[X_1,X_2]}^2(\ell)=&f_\mathrm{sky}(2\ell+1)\times\\&\frac{\left(C_\ell^{[X_1,X_2]}\right)^2}{\left[\tilde{C}_\ell^{[X_1,X_1]}\tilde{C}_\ell^{[X_2,X_2]}+\left(\tilde{C}_\ell^{[X_1,X_2]}\right)^2\right]}.
\end{aligned}
\end{equation}

The total SNR for the full spectrum across all bins, after considering the covariance, is given by
\begin{equation}\label{eq:28}
\mathrm{SNR}_{\mathrm{total}}^2(\ell)=f_{\mathrm{sky}}\left(2\ell+1\right)\mathbf{C}_\ell^\top\cdot\mathrm{Cov}^{-1}\cdot\mathbf{C}_\ell.
\end{equation}

Hence, the cumulative SNR can be obtained by summation:
\begin{equation}\label{eq:29}
\mathrm{SNR}^2(\ell<\ell_\mathrm{max})=\sum_{\ell^{\prime}=\ell_\mathrm{min}}^{\ell_\mathrm{max}}\mathrm{SNR}^2(\ell^{\prime}).
\end{equation}

\begin{figure*}
    \centering
    \includegraphics[width=1\linewidth]{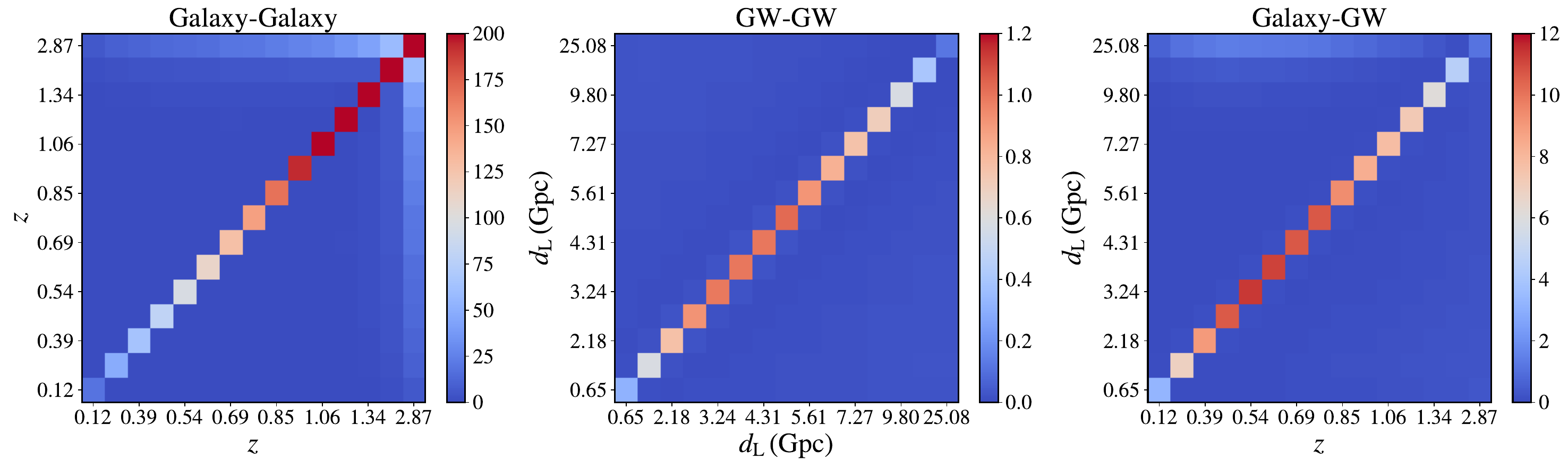}
    \caption{SNRs for the auto-correlation of the CSST photometric galaxy sample (left), the auto-correlation of GW source catalogs from the ET2CE detector network (center), and their cross-correlation (right). Each small square represents the analysis result for a specific combination of redshift and luminosity-distance bins, with the color indicating the SNR value within that bin.}
    \label{fig:snr_10yr}
\end{figure*}

Figure~\ref{fig:snr_10yr} visually presents SNRs of the auto-correlation and cross-correlation across different bins between the CSST photometric galaxies and the GW sources detected by the network composed of ET and two CEs (ET2CE). It also clearly demonstrates the effect described in eq.~\eqref{eq:11}, where the cross-correlation signal reaches its maximum in the redshift and luminosity distance bin that satisfies the correct distance-redshift relationship.  

\section{Modeling the characteristics of the surveys}\label{sec:simulation}

In this section, we describe the modeling of the observational characteristics for both the CSST photometric survey and the 3G ground-based GW detector network. Rather than generating actual mock catalogs, we adopt a parametric approach to characterize the expected source distributions and measurement uncertainties. Specifically, we outline the redshift distribution and redshift errors for the CSST galaxy samples, as well as the population distributions and parameter estimation errors for the GW events. These specifications serve as the foundational inputs for the cross-correlation analysis and cosmological Fisher forecasts.

\subsection{The galaxy catalog}\label{ssec:galaxy}

In this work, we consider the CSST photometric survey catalog. We select the CSST photometric catalog over the spectroscopic catalog primarily because the photometric survey offers a larger number of galaxies, which effectively suppresses shot noise and enhances statistical significance. Furthermore, it provides a broader redshift coverage, extending up to $z \sim 4$ \cite{CSST:2025ssq}, thereby maximizing the region that overlaps with GW events and strengthening the capability for cross-correlation detection.

The CSST photometric survey covers approximately 17500 deg$^2$. We adopt the galaxy number density distribution provided by ref.~\cite{Gong:2019yxt}, which includes about $1.925\times10^9$ galaxies, and account for a photometric redshift uncertainty of $0.05(1+z)$. In our calculation of angular power spectra, we assume that galaxies are isotropic in the sky and include a sky fraction factor $f_{\mathrm{sky}}$ to account for the sky coverage of the CSST photometric survey. To ensure a conservative and reliable analysis, we divide the redshift range of $z = 0$ to $z = 4$ into 15 equally populated redshift bins, serving as the baseline binning scheme. The left panel of Figure~\ref{fig:number_density} shows the overall redshift distribution as well as the distribution within individual bins after incorporating the redshift uncertainty.

\subsection{The GW event catalog}\label{ssec:GW}

We simulate GW source parameters based on the population models inferred from the first three observing runs of the LVK network. Specifically, we model the mass distribution of BBHs using the Power Law + Peak model \cite{Talbot:2018cva} and assume that the merger rate follows a form similar to that in ref.~\cite{Madau:2014bja}. We set the population hyperparameters as the median values of the posterior constraints of ref.~\cite{LIGOScientific:2021aug} using 42 BBH events, with the local merger rate set to $R(z=0)=23.9\ {\rm Gpc^{-3}\ yr^{-1}}$. 

To forecast the measurement precisions of 3G ground-based GW detectors, we employ the publicly available code \texttt{GWFish}\footnote{\url{https://github.com/janosch314/GWFish}} \cite{Dupletsa:2022scg} to simulate the instrumental errors of the source parameters of GW events. We also consider the luminosity distance errors arising from the peculiar velocities of galaxies and weak lensing. For specific equations, refer to ref.~\cite {Song:2022siz}. We calculate the SNRs of each GW event using \texttt{GWFish} and set the detection SNR threshold as $\rho_{\rm th}=8$. We implement additional filtering by requiring GW events with the relative luminosity distance errors $\Delta d_{\mathrm{L}} / d_{\mathrm{L}}$ smaller than 1. For ET2CE network, the total number of filtered GW event samples over one year is approximately $6.4\times10^4$. Based on the Planck 2018 cosmology, the 15 redshift bins are converted into 15 luminosity-distance bins, which are then applied to the analysis of the filtered GW samples.

\begin{figure*}
    \centering
    \includegraphics[width=0.4\linewidth]{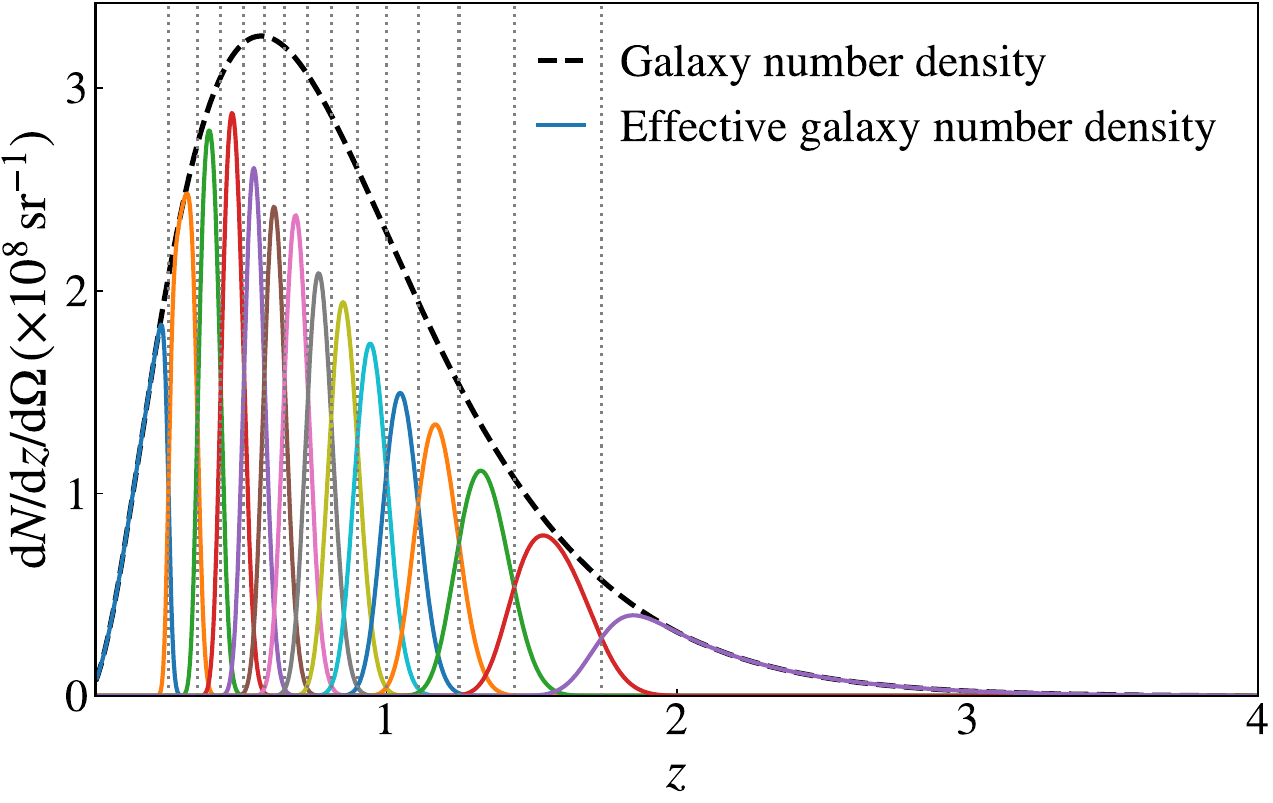}
    \hspace{0.3cm}
    \includegraphics[width=0.42\linewidth]{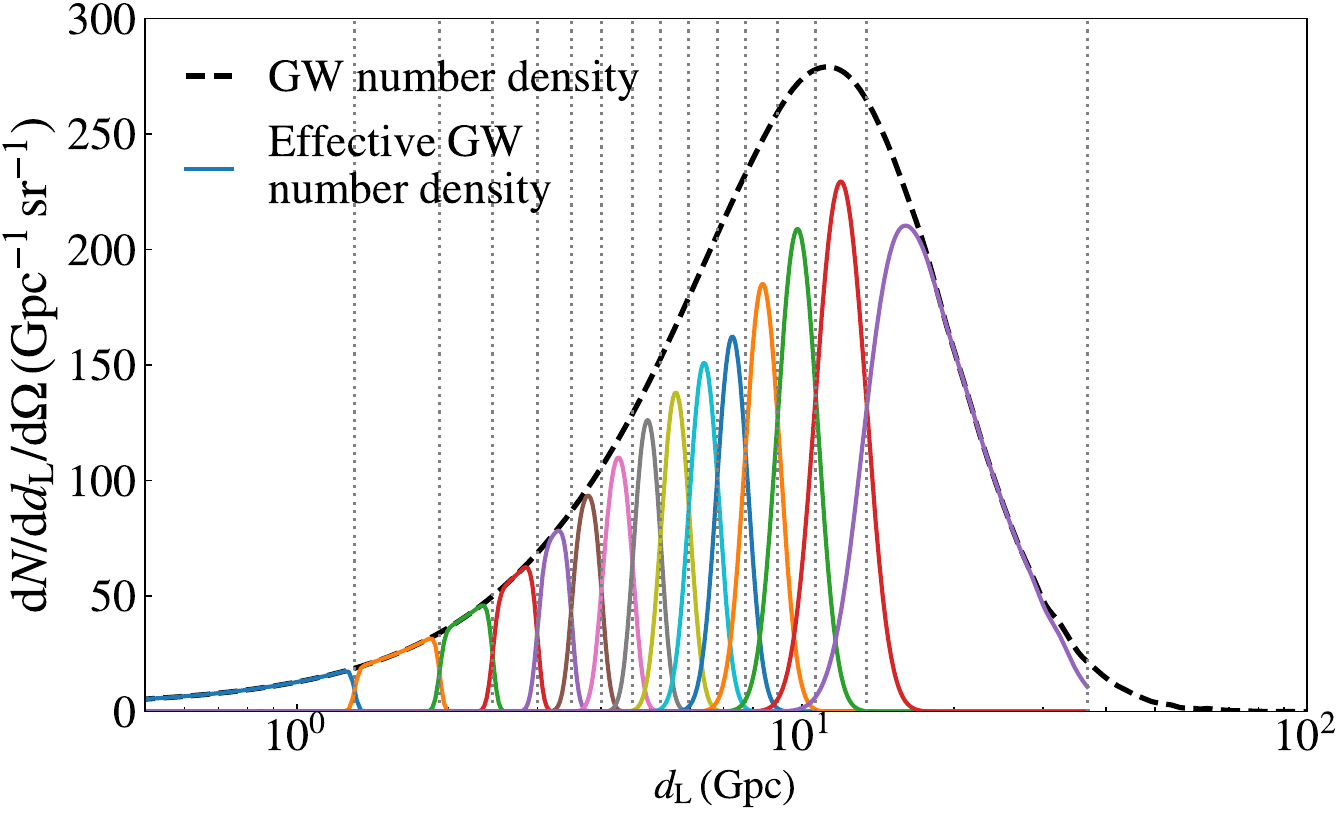}
    \caption{Number density distributions as a function of redshift. The left panel shows galaxies from the CSST survey, while the right panel presents GW events from 1 year by the ET2CE network. The dashed curves denote the original number distributions, and the solid curves indicate the effective number distributions after applying the window function as given in eq.~\eqref{eq:9}. Grey vertical lines mark the bin edges.} 
    \label{fig:number_density}
\end{figure*}

\begin{figure*}
    \centering
    \includegraphics[width=0.9\linewidth]{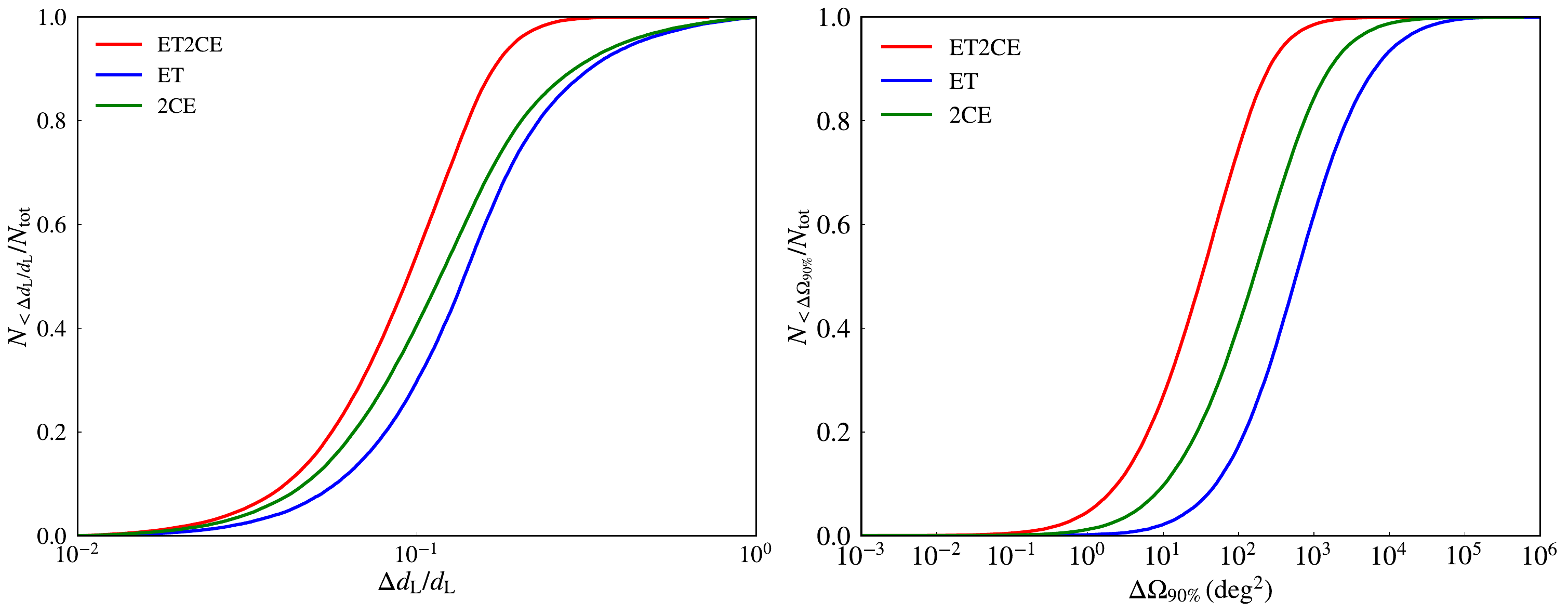}
    \caption{Cumulative distribution functions of the relative luminosity distance error $\Delta d_{\mathrm {L}}/d_{\mathrm {L}}$ and the 90\% sky localization error $\Delta \Omega_{90\%}$ for three GW detector configurations. The left panel presents $\Delta d_{\mathrm {L}}/d_{\mathrm {L}}$, where the y-axis value at a given x-axis value $X$ in the CDF curve represents the proportion of GW events with $\Delta d_{\mathrm {L}}/d_{\mathrm {L}}$ less than $X$ relative to the total number of events. The interpretation of the CDF curve for $\Delta \Omega_{90\%}$ in the right panel follows analogously.}
    \label{fig:CDF_errors}
\end{figure*}

The right panel of Figure~\ref{fig:number_density} shows the number density distributions of GW sources detected by the ET2CE detector network for a one-year observation period. The dashed curves denote the observed number density distributions, while the solid curves correspond to the effective number densities computed using eq.~\eqref{eq:9}. Figure~\ref{fig:CDF_errors} shows the parameter estimation uncertainties for GW sources with different detector configurations. The results demonstrate that the ET2CE network significantly improves both the luminosity distance and the sky localization measurement precisions, while also offering a substantial advantage in the number of detected events.

\begin{table*}
\centering
\caption{SNRs for different GW detector configurations, considering the CSST photometric catalog.}\label{tab:SNR}
\centering
\renewcommand{\arraystretch}{2}
\begin{tabular}{ccccccc}
\hline\hline 
\makebox[0.21\textwidth][c]{Configuration} & \makebox[0.21\textwidth][c]{GW auto SNR} & \makebox[0.21\textwidth][c]{Galaxy-GW cross SNR} & \makebox[0.21\textwidth][c]{Galaxy auto SNR}\\
\hline
ET2CE (1 yr) & 0.31 & 10.53 & --\\
ET2CE (5 yr) & 1.51 & 23.39 & --\\
ET2CE (10 yr) & 2.99 & 32.81 & --\\
ET (10 yr) & 0.72 & 8.87 & --\\
2CE (10 yr) & 1.70 & 19.68 & --\\
\textit{CSST galaxy sample only} & -- & -- & 1064.14\\

\hline\hline
\end{tabular}
\end{table*}

\section{Results and discussions}\label{sec:results}

In this section, we present the cross-correlation analysis between the 3G ground-based GW detectors and galaxy samples from the CSST 10-year photometric surveys. In the FIM analysis, the parameters varied include the cosmological parameters ($h$, $\Omega_{\rm c}$, $\Omega_{\rm b}$, $n_{\rm s}$, $A_{\rm s}$) and the GW clustering bias parameters ($A_{\mathrm{GW}}$, $\gamma$). The fiducial values of the cosmological parameters are set according to the Planck 2018 cosmology: $h = 0.6766$, $\Omega_{\rm c} = 0.26069$, $\Omega_{\rm b} = 0.04897$, $n_{\rm s} = 0.9665$, and $A_{\rm s} = 2.1\times10^{-9}$. For the primordial power spectrum amplitude, we adopt the parameter ${\rm ln}(A_{\rm s}\times10^{10})$ instead of $A_{\rm s}$ itself, because this logarithmic transformation can scale it to $\mathcal{O}(1)$, achieving a better numerical stability. However, it is worth noting that our current treatment of fixing the galaxy bias model $b_{\rm g}(z)$ may result in overly optimistic constraints on $A_{\rm s}$ due to their potential degeneracies. We will conduct more comprehensive analyses incorporating galaxy bias as free parameters in the future. The fiducial values for the GW clustering bias parameters are detailed in sect.~\ref{ssec:multi-tracer}.

We first calculate SNRs for the GW auto-correlation, the galaxy-GW cross-correlation, and the galaxy auto-correlation in different scenarios to explore the detectability of these power spectra, summarized in Table~\ref{tab:SNR}. We find that detecting the GW auto-correlation power spectrum may be challenging in the era of 3G ground-based GW detectors. When considering 10 years of BBH observations with ET, the SNR of the GW auto-correlation spectrum is only 0.72. Even with 10-year observations of the ET2CE network, the SNR increases only to 2.99. In contrast, the detection prospects for the galaxy-GW cross-correlation signal are much more promising. For example, considering ET with 10 years of observations and the CSST photometric redshift catalog, the cross-correlation SNR can reach 8.87. By comparison, the performance of the 2CE network shows a marked improvement, with a cross-correlation SNR reaching 19.68 after 10 years. With the ET2CE network, the detection capability is further enhanced: just 1 year of observations yields an SNR of 10.53, which increases to 32.81 after 10 years. This can be attributed to the higher effective number density of the ET2CE network, as shown in Figure~\ref{fig:number_density}, as well as its superior luminosity distance measurement precision and spatial localization accuracy, as illustrated in Figure~\ref{fig:CDF_errors}. In addition, we note that the galaxy auto-correlation SNR is as high as order $10^3$, reflecting the strong clustering signal and relatively low shot noise of the galaxy survey. 

\subsection{Constraints on cosmological parameters}\label{ssec:cosmological}

In this paper, we systematically examine the GW auto-correlation power spectrum, the GW-galaxy cross-correlation power spectrum, and the galaxy auto-correlation power spectrum, and their combined ability to constrain cosmological parameters. We also analyze the results for different GW detector configurations and observation durations. We summarize the results in Table~\ref{tab:results}.

We found that using the galaxy auto-correlation power spectrum alone can yield an $H_0$-constraint precision of about 5.53\%. In contrast, the GW auto-correlation power spectrum, with its significantly lower SNR compared to the galaxy auto-correlation, is far less effective in independently constraining cosmological parameters. Even with 10 years of observations from the ET2CE network, the constraint precision on $H_0$ only reaches 30.51\%, while other cosmological parameters remain entirely unconstrained. In the cross-correlation analysis between GW sources and galaxies, the constraints on cosmological parameters depend on the distance-redshift relation. Therefore, the cross-correlation can well constrain $H_0$ analogous to that obtained from GW standard sirens. For example, the cross-correlation spectrum of galaxies and the ET2CE 10-year GW events can constrain $H_0$ to about 1.06\%, which is an improvement of about 80.83\% compared with the results of the galaxy auto-correlation power spectrum.

Notably, combining the GW auto-correlation, the GW-galaxy cross-correlation, and the galaxy auto-correlation spectra further improves the constraint on cosmological parameters. For instance, In a joint analysis of three spectra using 10-year observations of ET2CE, the constraints on $H_0$ can reach 1.04\%, $\Omega_{\rm m}$ can reach 1.77\%, $n_{\rm s}$ can reach 1.73\%, and $A_{\rm s}$ can reach 2.76\%, improved by about 1.89\%, 66.92\%, 67.84\%, and 89.23\%, respectively, compared with GW-galaxy cross-correlation. This improvement arises from the distinct sensitivities of the galaxy auto-correlation and GW-galaxy cross-correlation to cosmological parameters. The galaxy auto-correlation constrains parameters through the clustering of galaxies, while the cross-correlation depends on the $d_{\rm L}$-$z$ relation that maximizes the correlation between galaxies and GW sources. Their different degeneracy directions help break parameter degeneracies in a joint analysis, thereby significantly enhancing the constraint on cosmological parameters. It is worth noting that in the joint constraint with three power spectra, the $H_0$ constraint mainly arose from the GW-galaxy cross-correlation.

\begin{figure*}
    \centering
    \includegraphics[width=1\linewidth]{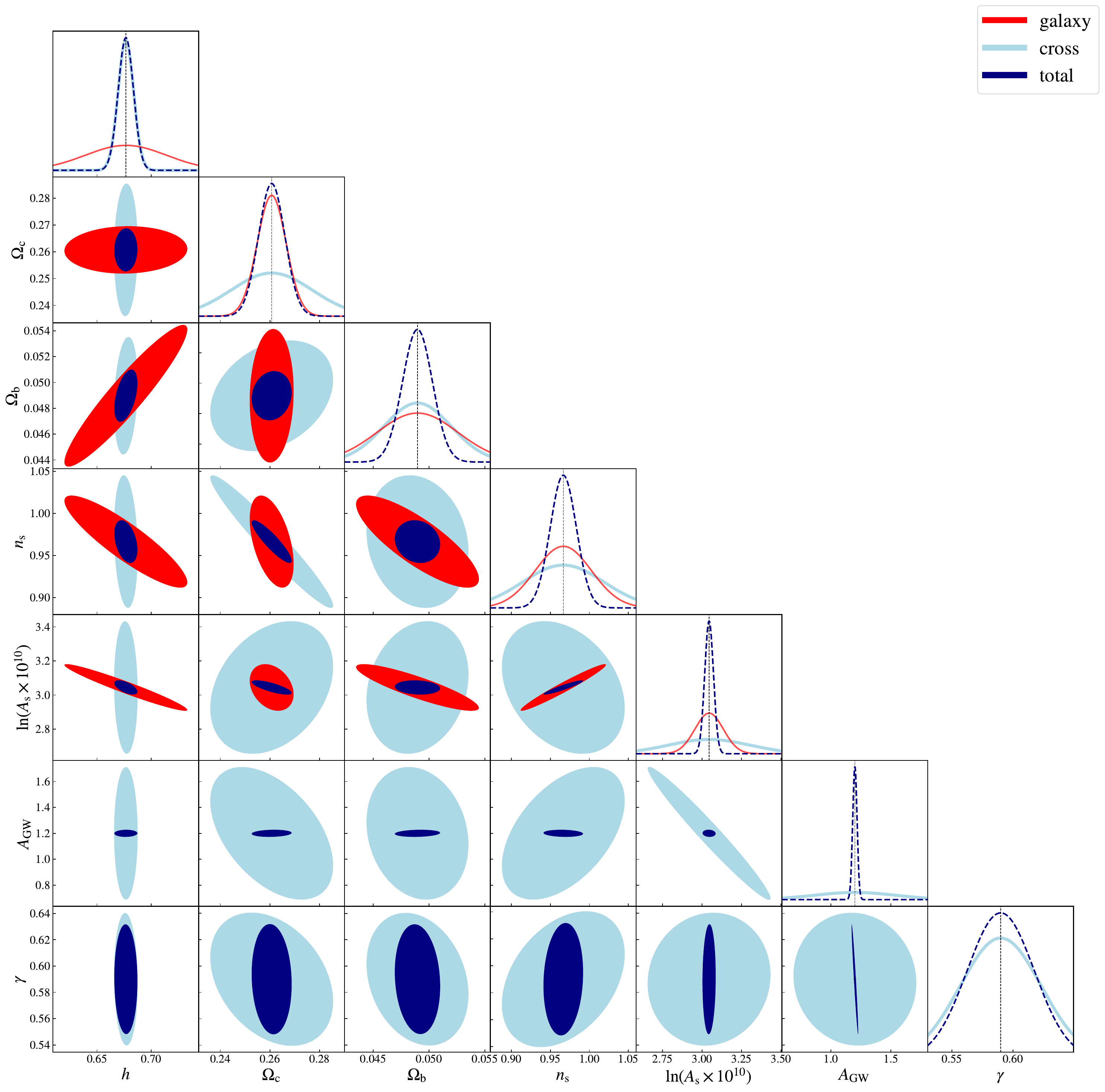}
    \caption{Constraints on cosmological and GW clustering bias parameters derived from the individual analyses of the galaxy auto-correlation, the galaxy-GW cross-correlation, and the joint analysis of both alongside the GW auto-correlation, respectively, using the 10-year GW sample of ET2CE and the CSST photometric galaxy sample.}
    \label{fig:10yr_3_corner}
\end{figure*}

In Figure~\ref{fig:10yr_3_corner}, we present the posterior distributions of cosmological parameters and the GW clustering bias parameters. It is evident that the galaxy auto-correlation and the galaxy-GW cross-correlation exhibit distinct degeneracy directions for almost all cosmological parameters, particularly between $H_0$ and $\Omega_{\rm c}$, where the degeneracy directions are nearly orthogonal. As a result, their combination effectively breaks parameter degeneracies and significantly improves parameter constraint precisions.

\begin{table*}
\centering
\caption{1$\sigma$ constraint precisions on cosmological and the GW clustering bias parameters from different power spectra for various GW detector configurations and observation times. All results are in percentages. The $\Omega_{\rm m}$ constraint precisions are derived from the $\Omega_{\rm c}$ and $\Omega_{\rm b}$ constraint precisions using the error propagation formula.}\label{tab:results}
\centering
\renewcommand{\arraystretch}{2}
\begin{tabular}{ccccccccc}
\hline\hline 
\makebox[0.09\textwidth][c]{GW data} & \makebox[0.09\textwidth][c]{$\Delta H_0/H_0$} &  
\makebox[0.09\textwidth][c]{$\Delta \Omega_{\rm c}/\Omega_{\rm c}$} & \makebox[0.09\textwidth][c]{$\Delta \Omega_{\rm b}/\Omega_{\rm b}$} & \makebox[0.09\textwidth][c]{$\Delta \Omega_{\rm m}/\Omega_{\rm m}$} & \makebox[0.09\textwidth][c]{$\Delta n_{\rm s}/n_{\rm s}$} & \makebox[0.09\textwidth][c]{$\Delta A_{\rm s}/A_{\rm s}$} & \makebox[0.09\textwidth][c]{$\Delta A_{\rm GW}/A_{\rm GW}$} & \makebox[0.09\textwidth][c]{$\Delta \gamma/\gamma$}\\
\hline
\multicolumn{9}{c}{\textbf{GW auto-spectra + Galaxy auto-spectra + Galaxy-GW cross-spectra}} \\

ET2CE (1 yr) & 2.18\% & 2.07\% & 3.62\%& 1.84\% & 2.12\% & 4.15\% & 1.52\% & 4.67\%\\
ET2CE (5 yr) & 1.44\% & 2.07\% & 2.98\% & 1.80\% & 1.86\% & 3.21\% & 1.52\% & 4.67\%\\
ET2CE (10 yr) & 1.04\% & 2.04\% & 2.73\% & 1.77\% & 1.73\% & 2.76\% & 1.52\% & 4.67\%\\

ET (10 yr) & 2.50\% & 2.18\% & 4.05\% & 1.95\% & 2.28\% & 4.61\% & 3.30\% & 13.42\%\\
2CE (10 yr) & 1.57\% & 2.11\% & 3.20\% & 1.85\% & 1.92\% & 3.39\% & 2.08\% & 7.35\%\\




\multicolumn{9}{c}{\textbf{Galaxy-GW cross-spectra}} \\

ET2CE (1 yr) & 2.39\% & 7.70\% & 6.96\% & 6.58\% & 6.73\% & 26.68\% & 30.08\% & 5.78\%\\
ET2CE (5 yr) & 1.50\% & 7.52\% & 6.42\% & 6.41\% & 6.40\% & 26.49\% & 29.83\% & 5.76\%\\
ET2CE (10 yr) & 1.06\% & 6.26\% & 6.14\% & 5.35\% & 5.38\% & 25.63\% & 28.12\% & 5.64\%\\

ET (10 yr) & 2.81\% & 13.29\% & 19.61\% & 11.61\% & 11.75\% & 89.69\% & 84.45\% & 22.77\%\\
2CE (10 yr) & 1.64\% & 7.18\% & 9.66\% & 6.24\% & 6.17\% & 39.62\% & 39.38\% & 9.81\%\\

\multicolumn{9}{c}{\textbf{Galaxy auto-spectra}} \\
-- & 5.53\% & 2.23\% & 7.40\% & 2.22\% & 3.73\% & 9.02\%& -- & --\\
                            
\hline\hline
\end{tabular}
\end{table*}

\subsection{Constraints on GW clustering bias parameters}\label{ssec:GW clustering bias}

In this work, we jointly constrain the cosmological parameters and the GW clustering bias parameters. In this subsection, we present the constraint results for the GW clustering bias parameters. Using the parameterization given in eq.~\eqref{eq:GW bias}, we infer the values of $A_{\mathrm{GW}}$ and $\gamma$, with the results summarized in Table~\ref{tab:results}. We find that the GW auto-correlation alone cannot effectively constrain the GW clustering bias parameters, while the GW-galaxy cross-correlation power spectrum provides significantly stronger constraints on the GW clustering bias. With 10-year observations of the ET2CE network, the constraint precisions on $A_{\mathrm{GW}}$ and $\gamma$ reach 28.12\% and 5.64\%, respectively. When combining the GW auto-correlation, GW-galaxy cross-correlation, and galaxy auto-correlation power spectra, the constraint precisions reach 1.52\% and 4.57\%, respectively, for 10-year ET2CE data. This improvement mainly arises from the enhanced constraints on cosmological parameters. As shown in Figure~\ref{fig:10yr_3_corner}, there exist degeneracies between the cosmological and the GW clustering bias parameters, and combining the three spectra helps break these degeneracies, thereby tightening the constraints on the GW clustering bias parameters.

Accurate constraints on the clustering bias parameters of GW sources are crucial for unveiling their origins and evolutionary pathways. Different types of GW sources, such as stellar-origin black holes and primordial black holes, exhibit fundamental differences in their formation mechanisms, environmental dependencies, and cosmic evolution histories, which are reflected in their distinct clustering biases. By analyzing the cross-correlations between GW sources and galaxies, one can effectively distinguish between these populations, providing key observational evidence for understanding the formation channels of GW sources and the evolution of cosmic structures \cite{Bosi:2023amu, Libanore:2023ovr}.

\begin{figure}
    \centering
    \includegraphics[width=1\linewidth]{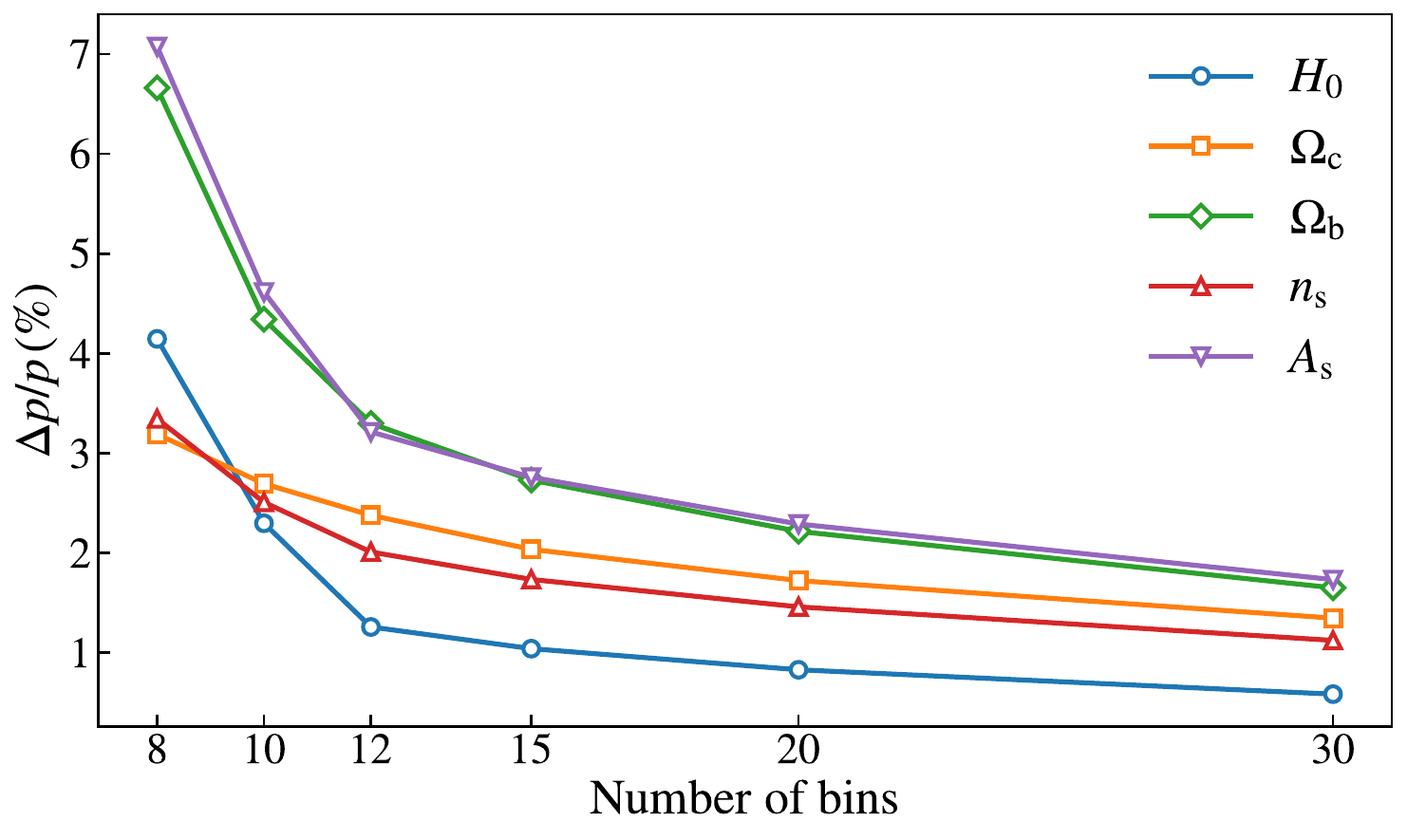}
    \caption{Constraint precisions of $H_0$, $\Omega_{\rm c}$, $\Omega_{\rm b}$, $n_{\rm s}$, and $A_{\rm s}$ as a function of the number of bins obtained by using the CSST photometric galaxy sample combined with the 10-year GW sample of ET2CE.}
    \label{fig:number_of_bins}
\end{figure}

\begin{figure}
    \centering
    \includegraphics[width=1\linewidth]{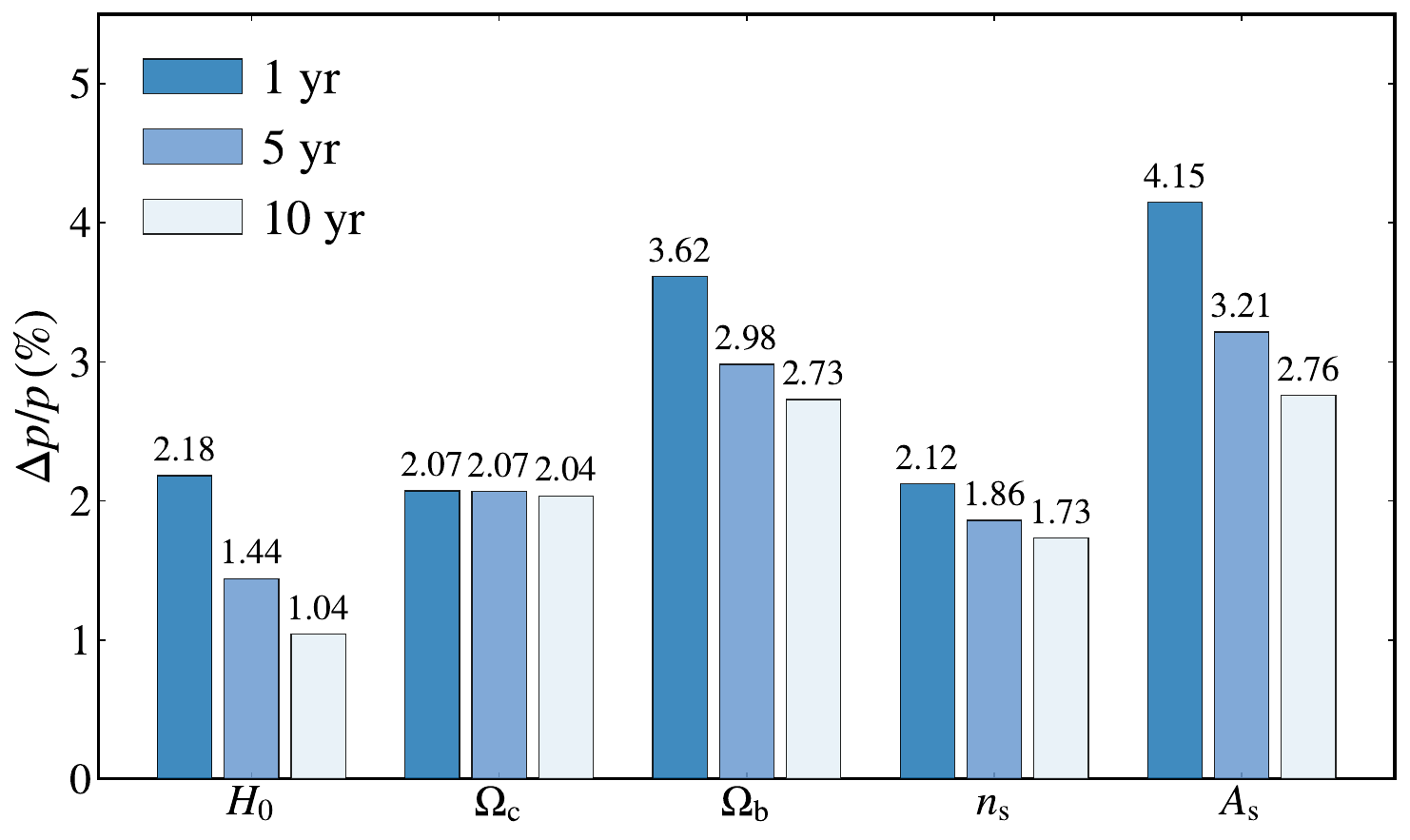}
    \caption{Constraint precisions of $H_0$, $\Omega_{\rm c}$, $\Omega_{\rm b}$, $n_{\rm s}$, and $A_{\rm s}$ obtained by using the CSST photometric galaxy sample combined with the GW sample of ET2CE, considering different GW observation durations.}  
    \label{fig:relative_uncertainty}
\end{figure}

\subsection{The factors affecting constraint precisions of cosmological parameters}\label{ssec:factors}

We further investigate the potential factors influencing the constraints on cosmological parameters. First, we discuss the impact of different GW detector configurations on the results. Compared with a single detector such as ET, network observations of 2CE and ET2CE provide significantly tighter constraints on both cosmological parameters and GW clustering bias parameters. 
For example, in the GW–galaxy cross-correlation analysis, using 10-year observations from the 2CE network compared with a single ET 10-year observation improves the constraint precisions on $H_0$, $\Omega_{\rm m}$, $n_{\rm s}$, $A_{\rm s}$, $A_{\rm GW}$, and $\gamma$ by 37.20\%, 5.13\%, 15.79\%, 26.46\%, 36.97\%, and 45.23\%, respectively. Furthermore, adopting 10-year ET2CE observations compared with 2CE further improves the constraint precisions on the same parameters by 33.76\%, 4.32\%, 9.90\%, 18.58\%, 26.92\%, and 36.46\%, respectively. This improvement primarily results from the enhanced sky localization precision of detector networks, which reduces the sky localization uncertainty $\Delta \Omega$ and enables access to higher multipoles $\ell_{\rm max}$. Consequently, more small-scale information can be incorporated into the analysis, leading to improved parameter constraints. In addition, GW network observations yield lower luminosity distance uncertainties $\Delta d_{\rm L}$, which affect the window function in eq.~\eqref{eq:9} and thus influence the effective number density of GW sources. 

We also explore various binning schemes to evaluate the robustness of our results, examining the impact of different bin counts as shown in Figure~\ref{fig:number_of_bins}. It can be seen that as the number of bins increases, the relative uncertainties of cosmological parameters generally decrease. However, when the number of bins exceeds approximately 20, the rate of improvement in parameter constraint precision slows noticeably. 

We further investigate the impact of the GW observation duration on the results. In Figure~\ref{fig:relative_uncertainty}, we show how different observation durations affect the constraint precision of cosmological parameters in the joint analysis of the CSST photometric sample and the ET2CE detector network. Our results on the trend of constraint precision improvement for parameters such as $H_0$ with observation duration are in excellent agreement with those shown in Figure 10 of ref.~\cite{Pedrotti:2025tfg}. The results indicate that longer observation times significantly improve the constraints on cosmological parameters. For instance, the relative uncertainty of $H_0$ decreases from 2.18\% for 1 year of observations to 1.04\% for 10 years. This improvement is primarily due to the increase in the number of detected GW events with longer observation durations, which effectively reduces shot noise and thus enhances the precision of cosmological parameter constraints.

\section{Conclusions}\label{sec:conclusions}

GW events and galaxies trace the same underlying matter distribution in the universe. By performing cross-correlation analyses between galaxies in different redshift bins and GW events in different luminosity-distance bins, one can establish a correspondence between the distance and redshift bins, thereby constraining cosmological parameters. Compared with traditional methods for inferring GW dark sirens' redshifts, this approach is less reliant on the completeness of galaxy catalogs and thus provides a complementary way for obtaining GW dark sirens' redshifts. In this work, we investigate the potential of cross-correlating CSST photometric galaxy samples with GW events from 3G ground-based GW detectors to constrain cosmological parameters.

Based on the population models inferred from GW data of the LVK's O1--O3 runs, we simulate GW sources, and then evaluate the GW source parameter estimation performance of 3G ground-based GW detectors using FIMs. We simulate galaxy samples using the galaxy distribution of the CSST photometric redshift survey given in ref.~\cite{Gong:2019yxt} and assume that the redshift measurement uncertainty is $\sigma_z(z)=0.05(1+z)$. We then compute the galaxy auto-correlation, the GW–galaxy cross-correlation, and the GW auto-correlation spectra. By using the FIM method, we evaluate the expected constraint precisions of cosmological parameters and GW clustering bias parameters using these power spectra and their combination.

The results demonstrate that the cross-correlation analysis between CSST photometric galaxies and GW events from 3G ground-based GW detectors can effectively constrain cosmological parameters and GW clustering bias parameters. Considering 10-year observations of ET2CE and combining the galaxy auto-correlation, the GW–galaxy cross-correlation, and the GW auto-correlation spectra, we constrain $H_0$, $\Omega_{\rm m}$, $n_{\rm s}$, and $A_{\rm s}$ to 1.04\%, 1.77\%, 1.73\%, and 2.76\%, respectively. When increasing the number of bins from 15 to 20, the constraint precisions of these cosmological parameters can reach 0.83\%, 1.49\%, 1.46\%, and 2.29\%, respectively. Furthermore, the joint analysis of the three angular power spectra can also constrain the GW clustering bias parameters $A_{\rm GW}$ and $\gamma$ to precisions of 1.52\% and 4.67\%, respectively.

We find that combining the CSST galaxy sample with GW samples from 3G ground-based GW detectors exhibits great potential for constraining both cosmological parameters and GW clustering bias parameters. (1) The deep redshift coverage of the CSST photometric survey effectively increases the number of analysis bins, enabling more precise parameter constraints. (2) The degeneracy directions of cosmological parameters in the galaxy-GW cross-correlation power spectrum differ from those in the galaxy auto-correlation spectrum, allowing the joint analysis to break parameter degeneracies and improve constraint precisions. (3) By jointly using the galaxy auto-correlation, GW-galaxy cross-correlation, and GW auto-correlation power spectra, the cosmological parameter constraint precisions can be further enhanced, which in turn improves the constraint precisions of GW clustering bias constraints, potentially offering valuable insights into the formation channels of GW sources.

\begin{acknowledgments}
We thank Furen Deng, Feng Shi, Zhihao Shang, and Diyang Liu for fruitful discussions. This work was supported by the National Natural Science Foundation of China (Grants Nos. 12473001, 12575049, 12533001, and 12305058), the National SKA Program of China (Grants Nos. 2022SKA0110200 and 2022SKA0110203), the China Manned Space Program (Grant No. CMS-CSST-2025-A02), the National 111 Project (Grant No. B16009), and the Natural Science Foundation of Hainan Province (Grant No. 424QN215).
\end{acknowledgments}

\section*{Data Availability}
The data that support the findings of this article are openly available \cite{du_2026_19201660}.

\bibliography{ref}

\end{document}